\documentclass[prd,a4paper,twocolumn, superscriptaddress,floatfix]{revtex4}
\usepackage{graphicx}
\usepackage{bbm}
\usepackage{amssymb}
\usepackage{hyperref}
\usepackage{upgreek}
\begin{document}
\newcommand{\be}{\begin{equation}}
\newcommand{\ee}{\end{equation}}
\newcommand{\bq}{\begin{eqnarray}}
\newcommand{\eq}{\end{eqnarray}}
\newcommand{\bsq}{\begin{subequations}}
\newcommand{\esq}{\end{subequations}}
\newcommand{\bc}{\begin{center}}
\newcommand{\ec}{\end{center}}
\newcommand\lapp{\mathrel{\rlap{\lower4pt\hbox{\hskip1pt$\sim$}} \raise1pt\hbox{$<$}}}
\newcommand\gapp{\mathrel{\rlap{\lower4pt\hbox{\hskip1pt$\sim$}} \raise1pt\hbox{$>$}}}
\newcommand{\dpar}[2]{\frac{\partial #1}{\partial #2}}
\newcommand{\sdp}[2]{\frac{\partial ^2 #1}{\partial #2 ^2}}
\newcommand{\dtot}[2]{\frac{d #1}{d #2}}
\newcommand{\sdt}[2]{\frac{d ^2 #1}{d #2 ^2}}
\newcommand{\ve}[1]{\mathbf{#1}}
\newcommand{\xd}{\dot{\ve{x}}}
\newcommand{\xpr}[1]{\ve{x}^{'\left(#1\right)}}
\newcommand{\xdh}{\dot{\rm{\hat{x}}}}
\newcommand{\xprh}[1]{\rm{\hat{x}}^{'\left(#1\right)}}
\newcommand{\xdhat}{\dot{\ve{\hat{x}}}}
\newcommand{\xprhat}[1]{\ve{\hat{x}}^{'\left(#1\right)}}
\newcommand{\vv}{\bar{v}}

\title{Cosmic Microwave Background anisotropies generated by domain wall networks}

\author{L. Sousa}
\email[Electronic address: ]{Lara.Sousa@astro.up.pt}
\affiliation{Instituto de Astrof\'{\i}sica e Ci\^encias do Espa{\c c}o, Universidade do Porto, CAUP, Rua das Estrelas, PT4150-762 Porto, Portugal}
\affiliation{Centro de Astrof\'{\i}sica da Universidade do Porto, Rua das Estrelas, PT4150-762 Porto, Portugal}

\author{P.P. Avelino}
\email[Electronic address: ]{pedro.avelino@astro.up.pt}
\affiliation{Instituto de Astrof\'{\i}sica e Ci\^encias do Espa{\c c}o, Universidade do Porto, CAUP, Rua das Estrelas, PT4150-762 Porto, Portugal}
\affiliation{Centro de Astrof\'{\i}sica da Universidade do Porto, Rua das Estrelas, PT4150-762 Porto, Portugal}
\affiliation{Departamento de F\'{\i}sica e Astronomia, Faculdade de Ci\^encias, Universidade do Porto, Rua do Campo Alegre 687, PT4169-007 Porto, Portugal}

\begin{abstract}
We develop a numerical tool for the fast computation of the temperature and polarization power spectra generated by domain wall networks, by extending the publicly available CMBACT code --- that calculates the CMB signatures generated by active sources --- to also describe domain wall networks. In order to achieve this, we adapt the Unconnected Segment model for cosmic strings to also describe domain wall networks, and use it to model the energy-momentum of domain wall networks throughout their cosmological history. We use this new tool to compute and study the TT, EE, TE and BB power spectra generated by standard domain wall networks, and derive a conservative constraint on the energy scale of the domain wall-forming phase transition of $\upeta <0.92\,\,{\rm MeV}$ (which is a slight improvement over the original Zel'dovich bound of $1\,\,{\rm MeV}$).
\end{abstract}

\maketitle

\section{Introduction\label{intro}}
Domain walls are two-dimensional topological defects that are formed when a discrete symmetry is spontaneously broken in a phase transition \cite{vilenkinbook}. The production of domain wall networks as remnants of phase transitions in the early universe is predicted in several grand unified scenarios. They are, however, often overlooked in cosmology since the energy scale of the phase transition that originates the domain walls is restricted to be smaller than  $1\,{\rm MeV}$ \cite{ZEL}. Nonetheless, there is still room for domain walls to play a relevant role in cosmology: for instance, they have been suggested as a possible significant dark energy contributor \cite{Bucher:1998mh}--- albeit they have since been ruled out as a major dark energy component \cite{PinaAvelino:2006ia,Avelino:2008ve,Sousa:2009is} -- and as a possible explanation for the spatial variations of the fine structure constant hinted at by HIRES/Keck and VLT/UVES data \cite{Olive:2010vh,Chiba:2011en,Avelino:2014xsa}.

Standard domain wall networks are expected to persist throughout cosmological history, and their presence at late times would necessarily leave imprints on the Cosmic Microwave Background (CMB). Although current CMB observations seem to be consistent with the inflationary paradigm, in which the fluctuations are seeded in the very early universe, they also allow for a subdominant topological defect contribution \cite{Ade:2013xla}. Domain walls, as cosmic strings, source metric perturbations actively throughout the cosmological history. For this reason, their CMB signatures are expected to be fundamentally different from those due to primordial fluctuations. In particular, cosmic defects are expected to generate a significant vector component that would not be present in inflation-seeded scenarios, because vector modes decay rapidly in the absence of a source \cite{Bardeen:1980kt}. The B-mode polarization signal originated by topological defects has then contributions from both tensor and vector modes, and may, for this reason, produce an observationally relevant signal in this channel, despite providing only subdominant contributions to the temperature and E-mode power spectra. The B-mode polarization channel offers thus a relevant observational window for the detection of topological defects.

Although some preliminary studies of the CMB signatures generated by domain walls have been done in the past \cite{ZEL,Turner:1990uw,Goetz:1990qz,Goetz:1990pj}), detailed studies of the anisotropies generated by these networks are yet to be performed. In this paper, we develop a numerical tool to compute the angular power spectrum generated by domain wall networks (in both the temperature and polarization channels), by adapting and extending the publicly available CMBACT code \cite{Pogosian:1999np,Pogosian:2006hg,cmbact} in order to accommodate these networks.

This paper is organized as follows. In Sec. \ref{dyn}, we briefly review the dynamics of infinitely thin and featureless domain walls. In Sec. \ref{model}, we adapt the Unconnected Segment Model (USM) to describe domain wall networks. We compute the energy-momentum tensor of a domain wall section in Sec. \ref{emt}. In Sec. \ref{res}, we present and discuss the temperature and polarization power spectra obtained using this method. We then conclude in Sec. \ref{conc}.

\section{Domain Wall Dynamics\label{dyn}}

The equation of motion of a domain wall is determined by the underlying field theory. However, if its thickness is negligible when compared to its curvature radii and if the domain wall is featureless --- as is often the case in cosmological scenarios ---, the domain wall may be treated as a 2-dimensional surface. While moving in spacetime, an infinitely thin and featureless domain wall sweeps an effectively $3$-dimensional worldvolume. The world history of a domain wall may thus be represented by

\be
x^{\mu}=x^{\mu}(\xi^a), \quad a=0,1,2\,,
\ee
where $\xi_0$ is a timelike parameter and $\xi_1$ and $\xi_2$ are spacelike parameters. These parameters may be regarded, at least locally, as coordinates on the worldvolume. The action of such a domain wall is given by \cite{vilenkinbook}

\be
S=-\sigma\int d^3\xi \sqrt{-h}\,,
\label{NGaction}
\ee
where $\sigma$ is the mass per unit area of the domain wall, $h=\det(h_{ab})$, and

\be 
h_{ab}=g_{\mu\nu}x^{\mu}_{,a}x^{\nu}_{,b}
\ee
is the metric induced on the worldvolume by the background (or pull-back metric).

By varying the Nambu-Goto action in Eq. (\ref{NGaction}) with respect to $x^{\mu}$, one obtains the following equation of motion for the dynamical variables $x^{\mu}$

\be
\frac{1}{\sqrt{-h}}\left(\sqrt{-h} h^{ab} x^{\mu}_{,b}\right)_{,a}+\Gamma^{\mu}_{\nu\lambda}h^{ab}x^{\nu}_{,a}x^{\lambda}_{,b}=0\,.
\label{eom1}
\ee

Let us now consider a domain wall in a flat Friedmann-Robertson-Walker (FRW) universe, whose line element is given by

\be
ds^2=a^2(\eta)\left[-d\eta^3+dr^2+r^2\left(d\theta^3+\sin^2\theta d^2\phi\right)\right]\,,
\ee
where $d\eta=dt/a(t)$ is the conformal time, $t$ is the physical time, $a$ is the cosmological scale factor, and $(r,\theta,\phi)$ are spherical coordinates.

The domain wall equation of motion should be invariant under re-parameterization of the worldvolume. In an expanding background, the temporal-tranverse gauge, with

\be
x^0=\eta\,,\qquad \mbox{and}\qquad \xd \cdot \xpr{i}=0\,,
\ee
would be a natural choice. Note that, since domain walls are assumed to be featureless, their physical velocity is purely orthogonal. Let us choose, for simplicity, a set of spacial parameters for the domain wall that are also orthogonal in the vicinity of the point under consideration, so that

\be
\xpr{i}\cdot\xpr{j}=0\,.
\ee
 Here, $x^{\mu}=\left(\eta,\mathbf{x}\right)$, dots and $'(i)$ represent derivatives with respect to conformal time and the $i$-th spatial parameter of the worldvolume, respectively, and the latin indices in parentheses refer to the spatial parameters of the wall.

In this gauge, Eq. (\ref{eom1}) yields

\bq
\ddot{\bf{x}}+3\mathcal{H}\left(1-\xd^2\right)\xd & = & \frac{1}{\epsilon}\left\{\left(\frac{\xpr{1}\xpr{2}}{\epsilon}\right)^{'(1)} + \right.\nonumber \\ & + & \left.\left(\frac{\xpr{1}\xpr{2}}{\epsilon}\right)^{'(2)}\right\}\label{eom2}\,,\\
\dot{\epsilon} & = & 3\mathcal{H}\epsilon \xd^2\,,\label{eom3}
\eq
where
\be
\epsilon=\sqrt{\frac{{\xpr{1}}^2 {\xpr{2}}^2}{1-\xd^2}}
\ee
is the coordinate energy per unit area, and $\mathcal{H}=\dot{a}/a$.

\section{Modeling the Domain Wall Network\label{model}}

The main objective of this article is to devise a numerical tool to compute the temperature and polarization power spectra generated by domain wall networks. In order to attain this goal, different approaches might be followed (as was done for cosmic strings): one may either use numerical simulations of domain wall networks, or use a phenomenological model to describe the network dynamics. Here, we choose to follow the latter option, given the adaptability of this model based approach: it allows for the description of different scenarios (by calibration of its free parameters), and it is not as constrained in terms of dynamical range as simulations often are.

For cosmic strings, this approach lead to the development of a publicly available tool for the computation of the temperature and polarization power spectrum generated by these networks --- the CMBACT code \cite{Pogosian:1999np,Pogosian:2006hg} --- that is based on the phenomenological Unconnected Segment Model (USM) \cite{Vincent:1996qr,Albrecht:1997mz,Battye:1997hu} to describe the energy-momentum tensor of the cosmic string networks. This tool was proven successful in predicting the shape of the angular power spectrum generated by cosmic string networks obtained using other models (see e. g. \cite{Copeland:1999gn,Bevis:2007qz}). It not only provides a good fit to the angular power spectrum predicted using Nambu-Goto simulations, but may also be calibrated in such a way as to mimic that obtained in Abelian-Higgs simulations (by calibrating it to match the field theory unequal time correlators predicted by field theory simulations) --- the so-called Abelian-Higgs mimic models described in \cite{Ade:2013xla}. Moreover, the bispectrum of cosmic string induced matter fluctuations predicted by CMBACT is in good agreement with that computed using perturbation theory \cite{Regan:2014vha}. The USM and CMBACT are, then, a flexible and robust framework to compute CMB anisotropies generated by cosmic strings, and, for this reason, we choose to extend them for domain wall networks.

\subsection{The Velocity-Dependent One-Scale Model for Domain Walls}

Let us now consider a domain wall network in a FRW universe, and assume that the network is statistically homogeneous on sufficiently large scales. In this case, two variables are sufficient to describe the large-scale cosmological evolution of the domain wall network. One such variable is the root-mean-squared (RMS) velocity of the network, $\vv=\sqrt{\left<v^2\right>}$, defined as

\be
\vv^2=\frac{\int v^2 \epsilon d^2 \xi}{\int \epsilon d^2\xi}\,,
\ee
where $v=\left|\xd \right|$ is the microscopic velocity. The other dynamical variable is the characteristic lengthscale of the network, $L$, defined as

\be
\rho=\frac{\sigma}{L}\,,
\label{den}
\ee
where $\rho$ is the average domain wall energy density.

The Velocity-dependent One-Scale (VOS) model provides a quantitative description of the large-scale dynamics of domain wall networks, by following the cosmological evolution of these two variables. The evolution equation for the RMS velocity may be obtained by averaging the microscopic equation of motion for a domain wall in Eq. (\ref{eom2}). On the other hand, one may obtain an equation of motion for the characteristic lengthscale of the network by differentiating the total energy density in domain walls --- given by

\be
E=\sigma a(\eta)\int\epsilon d^2 \xi\,,
\ee
 --- and using Eqs. (\ref{eom3})-(\ref{den}). These equations assume the form:
\bq
\dot{\vv} & = & \left(1-\vv^2\right)\left[\frac{\kappa}{l}-3\mathcal{H}\vv\right]\,,\\
\dot{l} & = & 3\vv^2\mathcal{H}l+\tilde{c}\vv\,,
\eq
and were derived in \cite{Avelino:2005kn,Sousa:2010zza,Avelino:2011ev}. (See also Refs. \cite{Sousa:2011ew,Sousa:2011iu} for a unified framework for the description of topological defects of arbitrary dimensionality.) Here, and for the rest of this paper, we choose to work with the comoving characteristic length, $l=L/a$. Moreover, we have introduced the phenomenological parameters $\tilde{c}$ and $\kappa$ that quantify, respectively, the energy loss caused by domain wall interactions and the effect of wall curvature on their dynamics. In this paper, we will assume these parameters take the values

\be
\tilde{c}=0.34\qquad\mbox{and}\qquad\kappa=0.98\,,
\ee
as indicated by the latest calibration of the VOS model against field theory simulations of standard domain wall networks \cite{Leite:2012vn}, during both the radiation and matter eras. Note that non-standard domain wall networks, such as networks with junctions, may also be describe by this model \cite{Avelino:2008ve}. In that case, however, these parameters would need to be recalibrated.

\subsection{An Unconnected Section Model for Domain Walls}

The VOS model merely provides a description of the large scale dynamics of a domain wall network. However, this is not sufficient to compute the cosmic microwave signature generated by domain walls: one also needs to characterize the energy-momentum tensor of the network in order to do so. To achieve this, we follow closely the USM for cosmic strings \cite{Vincent:1996qr,Albrecht:1997mz,Battye:1997hu}, and adapt it in order to describe domain walls. We preserve the essential elements of this framework and, therefore, we will only briefly review its essential features. In this model, the network of domain walls is represented by a collection of uncorrelated, flat and square domain walls, that have been produced simultaneously at some early time. The VOS model is used to set the comoving length of the network at any given time --- so that each wall has a comoving area $l^2(t)$ ---, and to fix the magnitude of the velocity of the segments.

The positions of the domain walls are drawn from an uniform distribution in space. Moreover, the direction of their velocity is chosen from a uniform distribution on a two sphere and, since we are assuming that the domain walls are featureless and, consequently, that their velocities are purely orthogonal, this direction also determines the orientation of the domain wall.

Throughout cosmic history, a fraction of domain walls decays at each epoch, so that the energy loss caused by domain wall interactions is taken into account. As is the case with the USM for cosmic strings, all domain walls that decay at the same time are consolidated in a single section. The number of sections that decay between instants $\eta_{i-1}$ and $\eta_i$ is given by

\be
\mathcal{N}(\eta_i)=\mathcal{V}\left[n\left(\eta_{i-1}\right)-n\left(\eta_{i}\right)\right]\,,
\ee
where $\mathcal{V}$ is the simulation volume, and $n\left(\eta\right)$ is the number density of domain walls

\be
n(\eta)=\frac{C\left(\eta\right)}{l^3\left(\eta\right)}\,.
\ee
Here $C\left(\eta\right)$ is determined by requiring that the number of walls is given by $\mathcal{V}/l^3\left(\eta\right)$ at any given time. In this way, at any given epoch, the number density of domain wall sections in the USM is in agreement with that predicted by the VOS model.

The decay of domain walls must necessarily be accompanied by a turning-off of the energy-momentum of that fraction. As is the case for the USM for cosmic strings, this effect is included by expressing the Fourier transform of the total energy-momentum as

\be
\Theta_{\mu\nu}\left(\bf{k},\eta\right)=\sum_{i=1}^{N}\left[\mathcal{N}(\eta_i)\right]^{\frac{1}{2}}\Theta^i_{\mu\nu} T^{\rm off}\left(\eta,\eta_i,L_{\rm f}\right)\,,
\ee
where the sum runs over all the $N$ domain wall sections, $\Theta^i_{\mu\nu}$ is the Fourier transform of the energy-momentum tensor of the $i$-th domain wall section. Furthermore, $T^{\rm off}\left(\eta,\eta_i,L_{\rm f}\right)$ is a function that controls domain wall decay and turns off their energy-momentum contribution:

\be
T^{\rm off}\left(\eta,\eta_i,L_{\rm f}\right)= \left\{
\begin{array}{cl}
1 \quad & \mbox{, for    $\eta<L_{\rm f}\eta_i$} \\
\frac{1}{2}+\frac{1}{4}\left(y^3-3y\right) \quad & \mbox{, for    $L_{\rm f}\eta_i<\eta<\eta_i$}\,,\\
0 \quad & \mbox{, for     $\eta_i<\eta$} \\
\end{array}
\right.
\ee
where

\be
y=2\frac{\ln(L_{\rm f}\eta_i)/\eta}{\ln(L_{\rm f})}-1\,,
\ee
and $L_{\rm f}$ is a parameter that controls the speed of domain wall decay.

\section{The Energy-Momentum Tensor of a Domain Wall\label{emt}}

With the USM  for domain wall networks set up, the only other ingredient missing is the energy-momentum tensor of the domain wall sections. The energy-momentum tensor of an infinitely thin and featureless domain wall may be obtained by varying the action in Eq. (\ref{NGaction}) with respect to the metric tensor $g_{\mu\nu}$:

\be
T_{\mu\nu}\sqrt{-g}=\sigma\int d^3\xi \delta^{(4)}\left[x^\mu-x^\mu(\xi^a)\right]\left\{\sqrt{-h}h^{ab}x^\mu_{,a}x^\nu_{,b}\right\}\,.
\ee

In the temporal-transverse gauge, it assumes the form
\begin{widetext}
\be
T_{\mu\nu}\left(\eta,\mathbf{x}\right)  =  \sigma \int d\eta d^2\xi\delta^{(4)}\left[x^{\mu}-x^{\mu}(\eta,\xi_1,\xi_2)\right]\left\{\epsilon{\dot{x}}^{\mu}{\dot{x}}^{\nu}-{\epsilon}^{-1}\left({\xpr{2}}^2{x^{'(1)}}^{\mu}{x^{'(1)}}^{\nu} + {\xpr{1}}^2 {x^{'(2)}}^{\mu}{x^{'(2)}}^{\nu}\right)\right\}\,,
\ee
\end{widetext}

where $\delta^{(4)}(x)$ is the 4-dimensional Dirac delta function.

Returning to the USM for domain walls, the Fourier transform of the energy-momentum tensor of a domain wall section is, then, given by
\begin{widetext}
\begin{eqnarray}
\Theta_{\mu\nu}(\ve{k},\eta) & = &\int d^3\mathbf{x} e^{i\ve{k}\cdot\ve{x}} T_{\mu\nu}\left(\eta,\xi_1,\xi_2\right)\\\nonumber
& = & \sigma\int_{-l/2}^{l/2}d\xi_1\int_{-l/2}^{l/2}d\xi_2 e^{i\ve{k}\cdot\ve{x}}\left\{\epsilon{\dot{x}}^{\mu}{\dot{x}}^{\nu}-
  {\epsilon}^{-1}\left({\xpr{2}}^2{x^{'(1)}}^{\mu}{x^{'(1)}}^{\nu} + {\xpr{1}}^2 {x^{'(2)}}^{\mu}{x^{'(2)}}^{\nu}\right)\right\}\,.
\end{eqnarray}
\end{widetext}

The spatial coordinates of a domain wall section may be expressed, in this gauge, as

\be
\ve{x}=\ve{x_0}+\xi_1\xprhat{1}+\xi_2 \xprhat{2}+v\eta \xdhat\,,
\ee
where $\ve{x_0}$ is the (random) position of its center of mass, $\xdhat$ and $\xprhat{i}$ are unitary vector with the direction of the velocity and of the spatial directions of the domain wall. Assuming, without loss of generality that $\ve{k}=k \bf{\hat{e}}_3$, we have that

\begin{widetext}
\bq
\Theta_{00} & = & 4\sigma\gamma\sqrt{2} \cos{\left(\ve{k}\cdot\ve{x}+vk\eta\xdh_3\right)} \frac{\sin{\left(kl{\xprh{1}}_3/2\right)}\sin{\left(kl{\xprh{2}}_3/2\right)}}{k^2{\xprh{1}}_3{\xprh{2}}_3}\,,\\
\Theta_{ij} & = & \Theta_{00}\left[v^2\xdh_i\xdh_j-\left(1-v^2\right)\left(\xprh{1}_i\xprh{1}_j+\xprh{2}_i\xprh{2}_j\right)\right]\,,
\eq
\end{widetext}
where the $\sqrt{2}$ factor was included to compensate for the fact that we are only considering the real part of $\Theta_{\mu\nu}(\ve{k},\eta$), and $\xdh_j=\xdhat \cdot \bf{\hat{e}} _j$ and $\xprh{i}_j=\xprhat{i} \cdot \bf{\hat{e}}_j$ are, respectively, the projections of vector $\xdhat$ and $\xprhat{i}$ along the $j$-th spatial direction (defined by the unitary vector $\bf{\hat{e}} _j$).

For this choice of $\bf{k}$, the Boltzmann integrator CMBFAST --- which is the basis of CMBACT -- requires five components of the energy-momentum tensor \cite{Albrecht:1997mz}. The first three are the scalar, vector and tensor components of the anisotropic stress, given by:

\bq
2\Theta_S & = & 2\Theta_{33}-\Theta_{11}-\Theta_{22}\,,\\
\Theta_V & = & \Theta_{13}\,,\\
\Theta_T & = & \Theta_{12}\,.
\eq
For single domain wall section, these yield 

\bq
2\Theta_S & = & \Theta_{00}\left\{v^2\left(3\xdh_3\xdh_3-1\right)-\right. \\
\nonumber & - & \left. \left(1-v^2\right)\left(3\xprh{1}_3\xprh{1}_3+3\xprh{2}_3\xprh{2}_3-2\right)\right\}\,,\\
\Theta_V & = & \Theta_{00}\left\{v^2 \xdh_1\xdh_3-\right.\\
\nonumber  & - & \left.  \left(1-v^2\right) \left(\xprh{1}_1\xprh{1}_3+\xprh{2}_1\xprh{2}_3\right)\right\}\,,\\
\Theta_T & = & \Theta_{00}\left\{v^2 \xdh_1\xdh_2-\right.\\ 
\nonumber  & - & \left. \left(1-v^2\right) \left(\xprh{1}_1\xprh{1}_2+\xprh{2}_1\xprh{2}_2\right)\right\}\,.
\eq
The remainder components --- the velocity field $\Theta_D$ (given by $\Theta_D= \Theta_{03}$ for this choice of $\bf{k}$), and the trace or isotropic pressure $\Theta=\Theta_{ii}$ --- may be obtained from the covariant conservation of energy-momentum:

\bq
\dot{\Theta}_D & = & -2\frac{\dot{a}}{a}\Theta_D-\nonumber \\
& - & \frac{k^2}{3}\left[\frac{a}{\dot{a}}\left(\Theta_D-\dot{\Theta}_{00}\right)-\Theta_{00}+2\Theta_S\right]=0\,,\\
\Theta & = & \frac{a}{\dot{a}}\left(\Theta_D-\dot{\Theta}_{00}\right)-\Theta_{00}\,.
\eq

\section{Results and Discussion\label{res}}

The CMB has a nearly perfect blackbody radiation spectrum \cite{Fixsen:1996nj}, with an approximately constant temperature accross the sky. For this reason, the basic CMB observables are its anisotropies, characterized by the temperature fluctuations

\be
\Delta (\ve{x}, \ve{n}, \eta_0)\equiv \frac{\left|T(\ve{x}, \ve{n}, \eta_0)-T_{\rm CMB}\right|}{T_{\rm CMB}}\,,
\ee
where $\bf{x}$ is the position of the observer, $\bf{n}$ is the line of sight direction, and $T_{\rm CMB}$ is the average temperature of the CMB. The contributions of the different angular scales may be separated by doing a decomposition into spherical harmonics

\be
\Delta (\ve{n})=\sum_{\ell m}a_{\ell m} Y_{\ell m} (\ve{n})\,,
\ee
where $Y_{\ell m}$ are spherical harmonic functions. The angular power spectrum, $C_{\ell}$, is defined as

\be
C_{\ell}\equiv \frac{1}{2\ell+1}\sum_{m=-1}^{\ell}\left\langle a_{\ell m}^*a_{\ell m}\right\rangle\,,
\ee
where angled brackets represent an ensemble average.

We calculate the CMB anisotropies generated by domain wall networks by implementing the changes described in Sec. \ref{emt} to the publicly available CMBACT code (version 4.0). This code is based on CMBFAST which integrates the Einstein and Boltzmann equations using the line of sight method  \cite{Hu:1997mn}. CMBACT also computes the the cold dark matter (CDM) linear power spectrum generated by active sources, 

\be
P(k)\equiv\left|\delta^2(\ve{k})\right|\,,
\ee
where $\delta(\ve{k})$ is the Fourier transform of the density contrast, 

\be
\delta(\ve{x})\equiv\frac{\rho(\ve{x})-\left\langle \rho\right\rangle}{\left\langle \rho\right\rangle}\,,
\ee
 $\rho(\ve{x})$ is the matter density at a given position $\ve{x}$, and  $\left\langle \rho\right\rangle$ is its average value.

Our results are found by averaging over $1000$ different realizations of a brownian domain wall network. We use the Planck 2015 cosmological parameters --- $\Omega_b^0 h^2=0.0225$ and $\Omega_m^0 h^2=0.1427$ for the baryon and matter density parameters, respectively, and $H_0=100h\,\,{\rm km s^{-1} {\rm Mpc}^{-1}}$, with $h=0.6727$, for the value of the Hubble parameter at the present time --- and fix the domain wall tension to $G\sigma L_0=10^{-7}$ (where $L_0$ is the characteristic lengthscale of the domain wall network at the present time).

\begin{figure}
\includegraphics[width=3in]{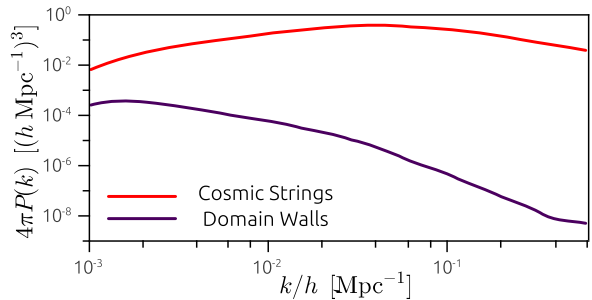}
\caption{Comparison between the linear CDM power spectrum generated by domain wall (purple line) and cosmic string (red line) networks. We have averaged over 1000 realizations of both string and wall networks, and chose $G\mu=G\sigma L_0=10^{-7}$.}
\label{ps}
\end{figure}

\begin{figure}
\includegraphics[width=3in]{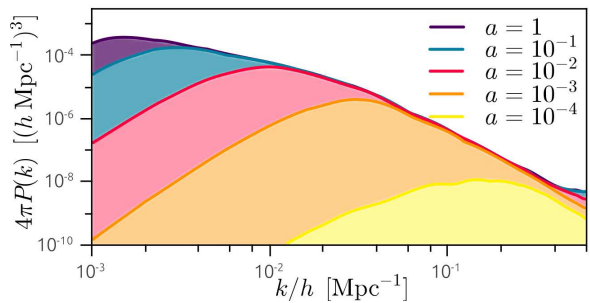}
\caption{Evolution of the linear CDM power spectrum generated by domain wall networks. We include the CDM power spectra generated by domain walls until $a=10^{-4}$ (yellow line), $a=10^{-3}$ (orange line), $a=10^{-2}$ (red line), $a=10^{-1}$ (blue line), and $a=1$ (purple line). We have averaged over 1000 realizations of domain wall networks, and chose $G\sigma L_0=10^{-7}$.}
\label{psevo}
\end{figure}

\begin{figure*}
\includegraphics[width=7.1in]{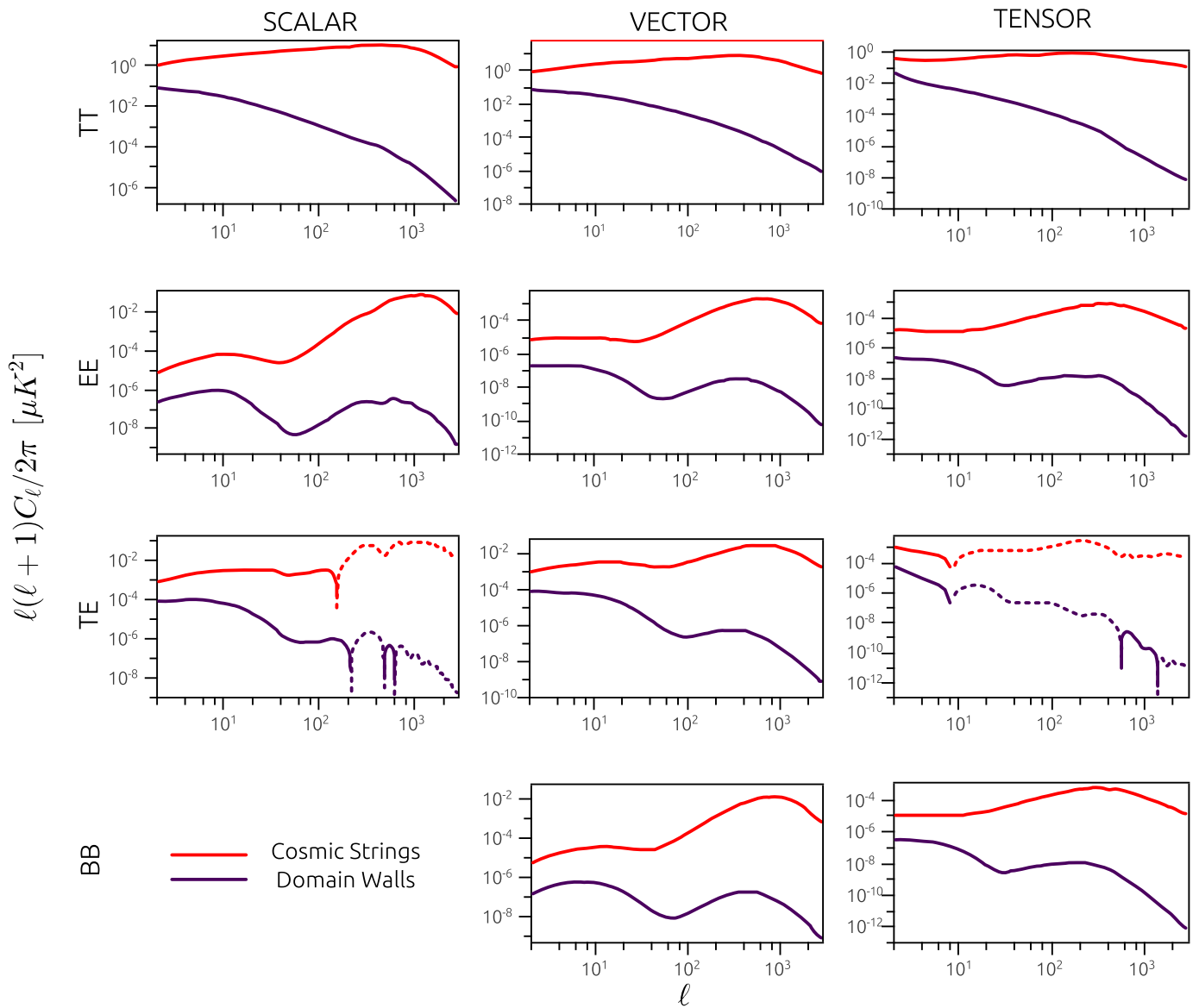}
\caption{Comparison between the angular power spectra generated by domain wall (purple lines) and by cosmic string (red lines) networks. From top to bottom, we plot the TT, EE, TE and BB power spectra, as a function of the multipole moment $\ell$. The left, middle and right panels represent the scalar, vector and tensor components, respectively. In the case of the TE power spectra, we chose to plot the absolute value and we use a dashed line to represent the parts in which $C_{\ell}$ is negative. We have averaged over 1000 realizations of both string and wall networks, and chose $G\mu=G\sigma L_0=10^{-7}$.}
\label{comb}
\end{figure*}

\begin{figure}
\includegraphics[width=3in]{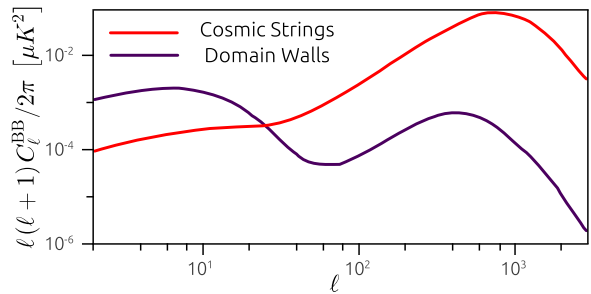}
\caption{Comparison between the total BB power spectra generated by domain wall (purple line) and cosmic string (red line) networks with the maximum tension allowed by current observational data. For domain walls, we have used the upper limit obtained in this paper $G\sigma L_0=5.6\times 10^{-6}$, while for cosmic string we took the upper limit obtained using Planck data $G\mu=2.4\times 10^{-7}$ \cite{Ade:2015xua}. We have averaged over 1000 realizations of both string and wall networks.}
\label{bmode}
\end{figure}

\begin{figure*}
\includegraphics[width=7.1in]{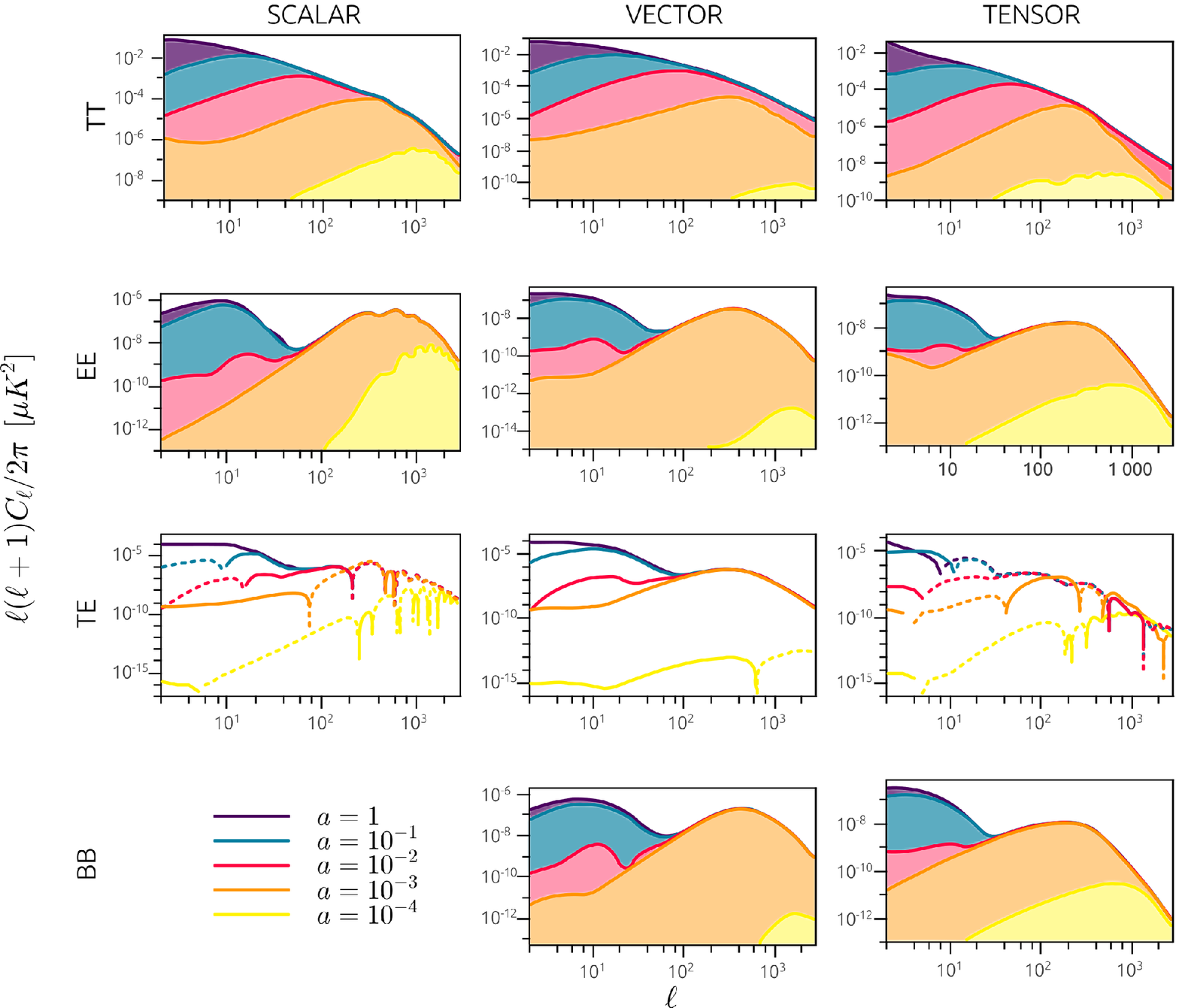}
\caption{Evolution of the angular power spectrum generated by domain wall networks. From top to bottom, we plot the TT, EE, TE and BB power spectra, as a function of the multipole moment $\ell$. The left, middle and right panels represent the scalar, vector and tensor components, respectively. In each of the plots we include the angular power spectra generated by domain walls until $a=10^{-4}$ (yellow line), $a=10^{-3}$ (orange line), $a=10^{-2}$ (red line), $a=10^{-1}$ (blue line), and $a=1$ (purple line). In the case of the TE power spectra, we chose to plot the absolute value and we use a dashed line to represent the parts in which $C_{\ell}$ is negative. We have averaged over 1000 realizations domain wall networks, and chose $G\sigma L_0=10^{-7}$.}
\label{combevo}
\end{figure*}

In Fig. \ref{ps}, we plot the CDM linear power spectrum generated by domain wall networks along that of cosmic string networks for $G\mu=G\sigma L_0=10^{-7}$ (we use the unaltered CMBACT to obtain the power spectra for cosmic strings). As this figure  illustrates, the matter power spectrum generated by domain walls is strongly suppressed on small scales (or large $k$), when compared to that of cosmic strings. Domain walls only become cosmologically relevant at late cosmological times, when the correlation length of the network is large. For this reason, these networks should be expected to contribute mostly to the matter power spectrum on large scales. Fig. \ref{psevo} --- where the evolution of the power spectrum generated by domain wall networks is plotted --- shows that this is indeed the case: the dominant contribution to the matter power spectrum at low $k$ is generated at late cosmological times, while the large $k$ contributions are mostly generated at earlier times (for $a<10^ {-1}$). Note however that, even on large scales, the CDM power spectrum generated by domain walls has an amplitude that is significantly smaller than that generated by cosmic strings. Domain walls have a correlation length that is slightly larger than that of cosmic strings at late times, which corresponds to a smaller energy density at the present time if $G\mu=G\sigma L_0$. This difference accounts for a factor of about 2 in the amplitude of the CDM power spectrum. However, the observed difference is slightly larger because cosmic strings become cosmologically relevant earlier than domain walls, thus contributing significantly to the CDM power spectrum at large scales over a longer period of time.

These  differences in the shapes of the matter power spectra generated by domain walls and cosmic strings must necessarily translate into significant differences in the CMB signatures of these defects. In Fig. \ref{comb}, we plot the TT, EE, TE and BB angular power spectra generated by domain wall networks along that generated by cosmic strings. First of all, the overall magnitude of the angular power spectra generated by domain wall networks on large angular scales (or small $\ell$) is lower than that of cosmic strings. Moreover, the strong suppression of the CDM power spectrum on small scales also results in a significantly reduced angular power spectra at large $\ell$ when compared to that of cosmic string networks. In particular, the peak that is observed in the cosmic string temperature power spectrum at intermediate scales is absent in the case domain walls. Unlike cosmic strings --- which exhibit the aforementioned peak around $\ell\sim 200$ --- domain walls contribute to the temperature power spectrum mostly on large scales. For domain walls, then, the multipole modes around $\ell=2$ have the highest constraining power on their fractional contribution to the observed temperature power spectrum. Note that the uncertainties at low $\ell$ are very large due to cosmic variance.  For this reason, as discussed in \cite{Bevis:2004wk}, observational data allows, on large angular scales, for a fractional contribution of cosmic strings to the TT power spectrum around ten times larger than the upper limit obtained using the full data set (which is more constraining mainly due to the peak located at intermediate angular scales, where the observational uncertainties are very small). Current Planck data allows for a fractional contribution of cosmic strings of about 1-2\% \cite{Ade:2013xla,Ade:2015xua}, and therefore one may conclude that the upper limit for the fractional contribution of cosmic strings in the TT channel using low $l$ data only should be around 10-20\%. One should expect the maximum contribution of domain walls to the TT power spectrum allowed by observational data to be similar to that allowed for cosmic strings at large scales. By taking the conservative approach and assuming the fractional contribution of domain walls to temperature power spectrum at $l=2$ to be around 20\%, we obtain a constraint on the domain wall mass per unit area of $\sigma<3.52\times 10^{-5}\,\,{\rm kg\,m^{-2}}$, which corresponds to a constraint on the domain wall-forming symmetry breaking scale of $\upeta<0.92\,\,{\rm MeV}$. This estimate, despite being conservative, is in good agreement with the Zel'dovich bound \cite{ZEL} --- which constrained $\upeta$ to be smaller than $1\,\,{\rm MeV}$ --- and even constitutes a slight improvement. Note however that the large cosmic variance associated to the CMB anisotropies generated by standard domain wall networks on large cosmological scales must necessarily be dealt with in a more rigorous study.

The aforementioned overall suppression of the temperature power spectra of domain walls (that becomes more accentuated with increasing $\ell$) is also observed in the polarization power spectra, and it is slightly more accentuated on large scales in the EE and BB channels. However, since the upper bounds on $G\sigma L_0$ are less stringent than those on $G\mu$, the contribution of domain walls to the polarization power spectrum at large scales may in fact be significant. In Fig. \ref{bmode}, we plot the total BB power spectrum generated by domain wall networks and cosmic string networks that have the maximum fractional contribution to the TT power spectrum allowed by current observational data. We chose $\sigma=3.52\times 10^{-5}\,\,\,{\rm kg\,m^{-2}}$ (or equivalently $G\sigma L_0=5.6 \times 10^{-6}$) for the domain wall networks, and the weakest constraint on cosmic string tension obtained using Planck data \cite{Ade:2015xua}, $G\mu=2.4\times 10^{-7}$ (which corresponds to the constraint on Abelian-Higgs strings). This figure shows that the B-mode polarization signal generated by domain wall networks that have a fractional contribution to the temperature power spectrum that is close to the allowed observational limit may dominate over that generated by cosmic strings (and the same is also true for the EE signal). The TT, EE and TE signals generated by topological defects are expected to be subdominant, when compared to that generated by primordial fluctuations. However, their contribution in the B-mode channel may be dominant, due to the presence of a vector component contribution that is absent in inflationary scenarios. Our results seem to indicate, therefore, that domain walls could be the dominant contributor in the BB channel for low multipole modes. Moreover, the fact that domain walls and cosmic strings have spectra with different shapes and that these defects contribute mostly at different scales should, in principle, allow to distinguish between these contributions if a signal is detected.

As is the case for the matter power spectrum, the dominant contributions to the CMB anisotropies are generated in the matter era. In Fig. \ref{combevo}, we plot the evolution of the TT, EE, TE and BB power spectra generated by a domain wall network. These plots show that domain walls contribute to the temperature anisotropies on progressively larger angular scales (or lower multipoles) as time progresses and their correlation length increases (as was the case for the CDM power spectrum). For the polarization power spectra, however, the picture that emerges is slightly different. Polarization may be created in two narrow windows in the history of the universe: very close to recombination and during the reionization epoch (because it requires the presence of a temperature quadrupole and the universe to be ionized). Fig. \ref{combevo} illustrates this fact: most of the contributions to the small scale peak are generated around the last scattering epoch (while $a\sim  10^{-3}$), while the large scale (dominant) peak is mostly generated at more recent cosmological times, around the time the scale factor was $a\sim10^{-1}$.

\section{Conclusions\label{conc}}

In this paper, we have expanded CMBACT to allow for the numerical computation of the CMB anisotropies generated by domain wall networks. This was done by adapting the USM for cosmic strings in order to also allow for the description domain wall networks, and by implementing the necessary changes on CMBACT (version 4.0). Note that, within this framework, the VOS model is used to set the correlation length and average velocity of the domain wall sections. This phenomenological model includes the essential aspects of domain wall dynamics, and it has the advantage of having two free parameters that may be used to calibrate the model in order to describe different types of domain wall networks. Our approach, thus, has the advantage of not being limited to a specific domain wall scenario.

We have also used this tool to study the angular power spectra generated by standard domain wall networks, in both the temperature and polarization channels, and to derive a conservative constraint on the energy scale of formation of the domain wall network of $\upeta<0.92\,\,{\rm MeV}$. Note, however, that this only applies to standard domain wall networks. Other domain wall scenarios --- such as networks whose dynamics is friction-dominated or domain wall networks with junctions --- would be characterized by a smaller correlation length. $L$ determines in which multipole mode the networks contribute dominantly to the temperature power spectrum. One should therefore expect constraints on non-standard networks to be stronger than the one obtained here, since they would be constrained by higher multipole modes which have smaller observational uncertainties and which are less affected by cosmic variance. The study of these different scenarios will be the subject of future work.

\textit{Note:} While this manuscript was in preparation, results related to the ones presented here appeared in Ref. \cite{Lazanu:2015fua}. The shape of the power spectra obtained in this work seems to be in good qualitative agreement with our results and the constraint on $\upeta$ presented by the authors is very similar to the one we have obtained. However, their constraint on $\sigma$  is four orders of magnitude smaller than the constraint we have obtained and it seems to be inconsistent with their upper limit of $\upeta$.
\acknowledgments

L.S. is supported by FCT through the grant SFRH/BPD/76324/2011. P.P.A. is supported by Funda{\c c}\~ao para a Ci\^encia e a Tecnologia (FCT, Portugal) FCT Research contracts of reference IF/00863/2012.  Funding of this work was also provided by the FCT grant UID/FIS/04434/2013.

\bibliography{dwcmb}

\end{document}